\documentclass[prd,eqsecnum,twocolumn,amsfonts,showpacs]{revtex4}

\usepackage{graphicx}

\usepackage{bm}

\newcommand{\beq}{\begin{equation}}
\newcommand{\eeq}{\end{equation}}
\newcommand{\beqs}{\begin{eqnarray}}
\newcommand{\eeqs}{\end{eqnarray}}

\begin{document}

\title{Further Study of an Approach to the Unification of Gauge Symmetries in
Theories with Dynamical Symmetry Breaking}

\author{Ning Chen}

\author{Robert Shrock}
%\thanks{robert.shrock@sunysb.edu}

\affiliation{
C.N. Yang Institute for Theoretical Physics \\
Stony Brook University, Stony Brook, NY 11794}

\begin{abstract}

We extend to larger unification groups an earlier study exploring the
possibility of unification of gauge symmetries in theories with dynamical
symmetry breaking. Based on our results, we comment on the outlook for 
models that seek to achieve this type of unification.

\end{abstract}


\maketitle

\section{Introduction}
\label{section1}

The origin of electroweak symmetry breaking is one of the most important
outstanding questions in particle physics.  One possibility is that this
breaking is caused by the formation of a bilinear condensate of new fermions
interacting via an asymptotically free, vectorial gauge interaction, called
technicolor (TC), that becomes strong at the TeV scale \cite{tc}. To
communicate the electroweak symmetry breaking to the quarks and leptons and
generate masses for these fermions, one embeds this theory in a larger one,
extended technicolor (ETC), containing gauge bosons that transform quarks and
leptons into the new fermions, and vice versa \cite{etc,etcrev}.  These
theories are subject to stringent constraints from precision electroweak
measurements and measurements of, or limits on, flavor-changing neutral
currents.  Modern theories of this type incorporate a gauge coupling that runs
slowly over an extended interval of energies to enhance quark and lepton
fermion masses.  Calculations indicate that this behavior can also reduce
technicolor corrections to the $Z$ and $W$ boson propagators somewhat
\cite{swtc,adss}; however, because of the strongly interacting nature of the
relevant physics, there remain significant theoretical uncertainties in the
estimates of these corrections.

  A natural question that arises in considering these theories with dynamical
electroweak symmetry breaking is how the technicolor gauge interaction might be
unified with the gauge group of the standard model (SM), $G_{SM}={\rm SU}(3)_c
\times {\rm SU}(2)_w \times {\rm U}(1)_Y$.  In Ref. \cite{lrs}, a partially
unified model of this type was constructed with the property that the electric
charge operator is a linear combination of generators of nonabelian gauge
groups, and hence electric charge is quantized.  Ideally, one would like to go
further and embed the TC gauge group $G_{TC}$, together with $G_{SM}$, in a
simple group, thereby relating the associated gauge couplings \cite{fs}.  In
Ref. \cite{tg} a study was carried out of several approaches to this type of
unification.

  Here we shall extend the analysis of Ref. \cite{tg}.  We consider models that
are designed to unify $G_{SM}$ with $G_{TC}$ or a larger gauge symmetry
described by a group $G_{SC} \supseteq G_{TC}$ (where $SC$ denotes ``strongly
coupled''), in a simple Lie group $G$,
\beq
G \supset G_{SC} \times G_{GU} \ . 
\label{scxgut}
\eeq
A notable feature of this approach is that it predicts the number of
generations of quarks and leptons, $N_{gen.}$.  A simple group $G_{GU}$ that
contains $G_{SM}$ has a lower bound on its rank of $rk(G_{GU}) \ge
rk(G_{SM})=4$, and the minimal nonabelian group that one could use for $G_{SC}$
has rank 2.  It follows that the rank of $G$ satisfies
\beq
rk(G) \ge rk(G_{SC})+rk(G_{GU}) \ge 6 \ .
\label{rkeq}
\eeq
It is natural to focus on SU($N$) groups, using ${\rm SU}(N_{SC}) \supseteq 
{\rm SU}(N_{TC})$ and
\beq
{\rm SU}(N) \supset {\rm SU}(N_{SC}) \times {\rm SU}(5)_{GU} \ , 
\label{embedding}
\eeq
where SU(5)$_{GU}$ is the usual grand unification group \cite{gg}, with
\beq
N = N_{SC} + 5 \ . 
\label{n}
\eeq
Since the group SU($N_{SC}$) involves interactions that should get strong at or
above the TeV scale, it must be asymptotically free and hence nonabelian. Since
the minimal value of $N_{SC}$ is thus 2, it follows that the minimal value of
$N$ is 7.  However, the $N=7$ case yields only two standard-model fermion
generations \cite{fs}. In Ref. \cite{tg}, cases up to $N=10$ were studied,
including a number that satisfy the requirement of yielding $N_g=3$
standard-model fermion generations, and some challenges for this unification
program were found.  Here we shall extend this study, considering the next two
higher cases, $N=11$ and $N=12$.  Based on our findings, we discuss aspects of
this approach to unification of theories with dynamical electroweak symmetry
breaking.

\section{General Structure of Unification Models}

We consider a general approach in which some SM fermion generations may arise
directly from the representations of the unified group $G$, while the remaining
ones arise indirectly, from sequential symmetry breaking of a subgroup of $G$
at ETC-type scales.  Let us denote $N_{gh}$ and $N_{g \ell}$ as the numbers of
standard-model fermion generations arising from these two sources,
respectively, where the subscripts $gh$ and $g \ell$ refer to
\underline{g}enerations from the representation content of the
\underline{h}igh-scale symmetry group and from the \underline{l}ower-scale
breaking.  The sum of these satisfies
\beq
N_{gen.} = 3 = N_{gh} + N_{g \ell} \ .
\label{ngen}
\eeq
At this stage the number $N_{g \ell}$ is only formal; that is, we construct a
model so that, {\it a priori}, it can have the possibility that a subgroup of
$G$ such as $G_{SC}$ might break in such a manner as to peel off $N_{g \ell}$
SM fermion generations.  However, we must examine for each model whether this
breaking actually occurs; this will be discussed further below. 

We next explain our procedure for analyzing the models; for further details,
the reader is referred to Ref. \cite{tg}.  The fermion representations are
determined by the structure of the fundamental representation, which we take to
be
\beq
\psi_R = \left( \begin{array}{c}
        (N^c)^\tau \\
         d^a \\
        -e^c \\
        \nu_e^c \end{array} \right )_R
\label{5genR}
\eeq
where $d$, $e$, and $\nu$ are generic symbols for the fermions with these
quantum numbers.  Thus, the indices on $\psi_R$ are ordered so that the indices
in the SC set, which we shall denote $\tau$, take on the values
$\tau=1,...N_{SC}$ and then the remaining five indices are those of the $5_R$
of SU(5)$_{GU}$, including the color index $a$ on $d^a$. The components of
$N^c_R$ transform according to the fundamental representation of SU($N_{SC}$),
are singlets under SU(3)$_c$ and SU(2)$_w$, and have zero weak hypercharge and
hence also zero electric charge. This structure is concordant with the direct
product in eq. (\ref{scxgut}) and the corresponding commutativity property
$[G_{SC}, G_{GU}]= 0$ and hence $[G_{TC}, G_{GU}]= 0$. (Recent discussions of
models with higher-dimensional representations of $G_{TC}$ include \cite{hr};
some other approaches to unification of $G_{TC}$ with SM gauge symmetries
include \cite{ou}.)

We next specify the fermion representations of $G={\rm SU}(N)$.  In the
following, we shall usually write the fermion fields as left-handed. In order
to avoid fermion representations of SU(3)$_c$ and SU(2)$_w$ other than those
experimentally observed, namely singlets and fundamental or conjugate
fundamental representations, we restrict the fermions to transform as $k$-fold
totally antisymmetrized products of the fundamental or conjugate fundamental
representation of SU($N$); these are denoted as $[k]_N$ and $[\bar
k]_N = \overline{[k]}_N$.  A set of (left-handed) fermions $\{ f \}$
transforming under $G$ is thus given by
\beq
\{ f \} = \sum_{k=1}^{N-1} \ n_k \ [k]_N
\label{fermionset}
\eeq
where $n_k$ denotes the multiplicity (number of copies) of each
representation $[k]_N$.  We use a compact vector notation
${\bf n} \equiv (n_1,...,n_{N-1})_N$. 
If $k=N-\ell$ is greater than the integral part of $N/2$, we shall work with
$[\bar \ell]_N$ rather than $[k]_N$; these are equivalent with respect to
SU($N$).

  An acceptable model should satisfy the following requirements: (i) the
contributions from various fermions to the total SU($N$) gauge anomaly must
cancel each other, yielding zero gauge anomaly; (ii) the resultant TC-singlet,
SM-nonsinglet left-handed fermions must comprise a well-defined set of
generations, i.e., must consist of $N_{gen.}=3$ copies of $[(1,\bar
5)_L+(1,10)_L]$, where the first number in parentheses signifies that these are
singlets under $G_{TC}$ and the second number denotes the dimension of the
SU(5)$_{GU}$ representation; and (iii) in order to account for neutrino masses,
one needs to have TC-singlet, electroweak-singlet neutrinos to produce Majorana
neutrino mass terms that can drive an appropriate seesaw \cite{nt,ckm}. Here 
these are also singlets under $SU(5)_{GU}$. 

  As another requirement, (v), the ETC gauge bosons should have appropriate
masses, in the range from a few TeV to $10^3$ TeV, so as to produce acceptable
SM fermion masses. This requirement cannot be satisfied if $G$ breaks directly
to the direct product group $G_{TC} \times G_{SM}$ at the unification scale
$M_{GU}$ as in early approaches to TC unification \cite{fr}.  The requirement
could be satisfied if the breaking of $G$ at $M_{GU}$ would leave an invariant
subgroup ${\rm SU}(2)_w \times G_{SCC}$, where
\beq
{\rm SU}(N_{SCC}) \supset {\rm SU}(N_{SC}) \times {\rm SU}(3)_c 
\label{gscc}
\eeq
with
\beq
N_{SCC} = N_{SC} + N_c = N_{SC}+3 \ . 
\label{nscc}
\eeq
Here SCC stands for the the SC group together with the color group.  As the
energy scale decreases, this intermediate symmetry $G_{SCC}$ should break at
ETC scales, evantually yielding the residual exact symmetry group ${\rm
SU}(2)_{TC} \times {\rm SU}(3)_c$. This can occur naturally if the SCC gauge
interaction is chiral and asymptotically free; as the energy scale decreases
and the SCC gauge coupling increases, it can thus trigger the formation of a
fermion condensate which self-breaks $G_{SCC}$.  This type of process in which
a strongly coupled chiral gauge interaction self-breaks via formation of a
fermion condensate has been termed ``tumbling'' \cite{rds}.  Further
requirements are that (vi) if $N_{SC} > N_{TC}$, there should be a mechanism to
break SU($N_{SC}$) to SU($N_{TC}$); (vii) the TC interaction should be
vectorial and asymptotically free, so that the TC gauge coupling gets large as
the energy scale decreases to the TeV scale, triggering the formation of a
technifermion condensate for EWSB; and (viii) the residual SU(3)$_c$ color
group should be asymptotically free.

Let us define a $(N-1)$-dimensional vector whose components are the values of
the anomaly $A([k]_N)$ with respect to SU($N$), ${\bf a} =
(A([1]_N),...,A([N-1]_N))$.  Then the constraint that there be no $G$ gauge
anomaly is the condition
\beq
{\bf n} \cdot {\bf a} = 0 \ .
\label{n.a}
\eeq
This is a diophantine equation for the components of the vector of
multiplicities ${\bf n}$, subject to the constraint that the components $n_k$
are non-negative integers (as well as additional constraints discussed below). 

It is convenient to display the transformation property of a fermion
representation of $G$ with respect to the subgroups $G_{SC}$ and SU(5)$_{GU}$
by the notation $({\cal R}_{SC},{\cal R}_{GU})$.  The number of (left-handed)
fermions that transform as singlets under $G_{SC}$ and $\bar 5$'s of
SU(5)$_{GU}$ is
\beq
N_{(1,\bar 5)} = n_{_{N_{SC}+4}}+n_4
\label{n15bar}
\eeq
and the number of (left-handed) fermions that transform as singlets under 
$G_{SC}$ and $10$'s of SU(5)$_{GU}$ is 
\beq
N_{(1,10)} = n_2 + n_{_{N_{SC}+2}} \ .
\label{n10}
\eeq
Hence, the requirement that the left-handed SC-singlet, SM-nonsinglet
fermions comprise equal numbers of $(1,\bar 5)$ and (1,10)'s implies the
condition
\beq
n_{_{N_{SC}+4}} + n_4 = n_2 + n_{_{N_{SC}+2}} \ . 
\label{generationcondition}
\eeq
The number of SM fermion generations $N_{gh}$ produced by the
representations of $G$ is given by either side of this equation;
\beq
N_{gh} = n_2 + n_{_{N_{SC}+2}} \ .
\label{ngenh}
\eeq
The remaining $N_{g \ell}$ generations of SM fermions arise via the breaking 
of $G_{SC}$.  
 electroweak-singlet neutrinos, arise, in general, from two
sources: (i) $[N_{SC}]_N$, when all of the $N_{SC}$ indices take values in
SU($N_{SC}$); and (ii) $[5]_N$, when all of the indices take values in
SU(5)$_{GU}$.  In the special case $N_{SC}=5$, these each contribute. Hence,
\beq
N_{(1,1)}=n_{_{N_{SC}}}+n_5 \ .
\label{n11}
\eeq
Electroweak-singlet neutrinos arise from fermions that are singlets under both
$G_{SC}$ and SU(5)$_{GU}$; there are $N_{(1,1)}=n_{_{N_{SC}}}+n_5$ of these.

With the envisioned sequential breaking of $G_{SCC}$ and 
$G_{SC}$ that would produce the $N_{g \ell}$ SM fermion generations,
one has $N_{g \ell} = N_{SCC} - (N_{TC}+N_c)$ and 
\beq
N_{g \ell} = N_{SC} - N_{TC} \ . 
\label{nglrel}
\eeq
The requirement that there be no (left-handed) fermions transforming
as singlets under SU($N_{SC}$) and in an exotic manner, as 5's or
$\overline{10}$'s of SU(5)$_{GU}$ is satisfied if
\beq
n_1 = 0, \quad n_{_{N_{SC}+1}}=0 
\label{no5}
\eeq
and
\beq
n_3 = 0, \quad n_{_{N_{SC}+3}}=0 
\label{no10bar}
\eeq
respectively.

\section{$N_{SC}=6$, $G={\rm SU}(11)$}

We next proceed to analyze the new models, and first consider the case where
$N_{SC}=6$, so that $N=N_{SC}+5=11$ and ${\bf n} = (n_1,...,n_{10})_{11}$.
With $N_{gh}+N_{g\ell}=N_{gen.}=3$ and $N_{SC}-N_{TC}=N_{g\ell}$, one has, {\it
a priori}, four possibilities for the manner in which the SM fermion
generations arise, as specified by $(N_{gh},N_{g \ell},N_{TC})$, namely
(3,0,6), (2,1,5), (1,2,4), and (0,3,3).  However, as we shall show, only the
cases with $N_{gh}=0$ and $N_{gh}=2$ are actually allowed by the various
constraints.  This SU(11) model was not studied in Ref. \cite{tg} because it
does not allow one to use the preferred, minimal value, $N_{TC}=2$.  This
latter value is preferred in order to minimize technicolor corrections to
precisely measured electroweak quantities and because it makes possible a
mechanism to produce light neutrino masses \cite{nt,ckm,lrs}.  However, if one
takes into account the fact that quasi-conformal behavior in the technicolor
theory can reduce the technicolor corrections to the $Z$ and $W$ boson
propagators, the effect of the larger value of $N_{TC}$ might not be too
serious.  The conditions (\ref{no5}) and (\ref{no10bar}) that the theory should
not contain any $5_L$ or $\overline{10}_L$ yield
\beq
n_1=n_3=n_7=n_9=0 \ , 
\label{nzerosu11}
\eeq
and eq. (\ref{generationcondition}) is 
\beq
N_{gh}=n_2+n_8=n_4+n_{10} \ . 
\label{ngenhsu11}
\eeq
The condition of zero gauge anomaly, eq. (\ref{n.a}), is 
\beq
7(n_2+4n_4+2n_5-2n_6)-20n_8-n_{10}= 0 \ . 
\label{anomalysu11}
\eeq
For a given value of $N_{gh}=3-N_{g \ell}$, these are three nondegenerate
linear equations for the six quantities $n_2$, $n_4$, $n_5$, $n_6$, $n_8$, 
and $n_{10}$. The solution entails the relation
\beq
n_5 = n_6+\frac{1}{14}(27n_8+29n_{10})-\frac{5}{2}N_{gh} \ . 
\label{n5su11}
\eeq
A necessary condition for an acceptable solution is thus that
\beq
27n_8+29n_{10}-35N_{gh}=0 \quad {\rm mod} \  14 \ .  
\label{su11necrel}
\eeq
Let $r$ be a non-negative integer.  We find two classes of such solutions: (i)
$N_{gh}=0$, $n_8=n_{10}=r$ and hence, from eq. (\ref{n5su11}), $n_5=n_6+4r$;
(ii) $N_{gh}=2$, $n_8=n_{10}=r$, and hence $n_5=n_6+4r-5$.  

We first consider solutions of class (i).  These have $N_{g\ell}=3$ and
$N_{TC}=3$.  Now $N_{gh}=n_2+n_8=n_4+n_{10}=0$, which implies that $r=0$,
$n_2=n_8=n_4=n_{10}=0$, and $n_5=n_6=s$, where $s$ is some positive
integer. The resultant vector ${\bf n}$ is
\beq
{\rm class} \ (i): \quad {\bf n} = (0,0,0,0,s,s,0,0,0,0) \ . 
\label{nsu11_sol_1}
\eeq
The minimal choice would be $s=1$, but for generality, we shall keep $s$
arbitrary.  Since $[6]_{11} \approx [\bar 5]_{11}$, this SU(11) theory has
left-handed chiral fermion content
\beq
s \{ [5]_{11}+[\bar 5]_{11} \}
\label{sfi}
\eeq
and thus is vectorial.  Consequently, the fermion content with respect to the
subgroups SU(9)$_{SCC}$ and SU(6)$_{SC}$ is also vectorial.
With respect to the subgroup
\beq
{\rm SU}(2)_w \times {\rm SU}(9)_{SCC} \ , 
\label{su2su9}
\eeq
the $[5]_{11}$ representation transforms as 
\beq
[5]_{11} = (1,[\bar 4]_9) + (2,[4]_9) + (1,[3]_9) \ , 
\label{5_11decomp29}
\eeq
where we use the $[k]_9$ notation for the representations of SU(9)$_{SCC}$ and
the well-known dimensions to label the representations of SU(2)$_w$. The
total fermion content with respect to the subgroup (\ref{su2su9}) is comprised
of $s$ copies of eq. (\ref{5_11decomp29}) and its conjugate.  We recall the
requirement that the SCC and SC interactions should be asymptotically free.
For a given gauge group $G_j$ with gauge coupling $g_j$ and 
$\alpha_j =g_j^2/(4\pi)$, the evolution of the gauge couplings 
as a function of the momentum scale $\mu$ is given by the 
beta function $\beta_j = d \alpha_j/dt = - b_0^{G_j}\alpha_j^2/(2\pi) +
O(\alpha_j^3)$. where $t=\ln \mu$. We find that the SU(9)$_{SCC}$
gauge interaction is non-asymptotically free. Here and below, for comparative
purposes, it will be useful to give the actual coefficients.  We calculate 
\beq
b_0^{SU(9)_{SCC}} = 3(11-28s) \quad ({\rm class} \ i ) \ , 
\label{b0su9}
\eeq
which is negative for any value $s \ge 1$. With respect to the subgroup
\beq
{\rm SU}(6)_{SC} \times {\rm SU}(5)_{GU} \ , 
\label{su6su5}
\eeq
the $[5]_{11}$ representation transforms as 
\beqs
[5]_{11} & = & (1,1) + ([1]_6, \bar 5) + ([2]_6,\overline{10}) + 
([3]_6,10) \cr\cr
& + & ([\bar 2]_6,5) + ([\bar 1]_6,1) \ , 
\label{5_11decomp}
\eeqs
where, aside from the overall singlet (1,1), we use the $[k]_6$ notation for
the representations of SU(6)$_{SC}$ and the well-known dimensions to label the
representations of SU(5)$_{GU}$. The fermion content of this model with
respect to the subgroup (\ref{su6su5}) is the sum of $s$ copies of
eq. (\ref{5_11decomp}) and its conjugate.  The SU(6)$_{SC}$ gauge interaction
is not asymptotically free; the leading coefficient of its beta function is
\beq
b^{SU(6)_{SC}}_0=2(11-42s) \quad ({\rm class} \ i ) \ , 
\label{b0su6}
\eeq
which is negative for any $s \ge 1$.  This disfavors the model. 

We next consider models of class (ii).  These have $N_{g\ell}=1$ and
$N_{TC}=5$. The relations $N_{gh}=n_2+n_8=n_4+n_{10}=2$, together with the
assignment $n_8=n_{10}=r$ imply that
\beq
n_2=n_4=2-r \ . 
\label{caseiin2n4rel}
\eeq
We thus have three subclasses of solutions, namely
(ii.a) $r=2$, whence $n_2=n_4=0$ and $n_5=n_6+3$; 
(ii.b) $r=1$, whence $n_2=n_4=1$ and $n_5=n_6-1$; and
(ii.c) $r=0$, whence $n_2=n_4=2$ and $n_5=n_6-5$. 
Minimal choices in each of these three subclasses have the following ${\bf n}$
vectors:
\beq
(iia): \quad {\bf n} = (0,0,0,0,3,0,0,2,0,2)
\label{nsu11_sol_2a}
\eeq
\beq
(iib): \quad {\bf n} = (0,1,0,1,0,1,0,1,0,1)
\label{nsu11_sol_2b}
\eeq
\beq
(iic): \quad {\bf n} = (0,2,0,2,0,5,0,0,0,0) \ . 
\label{nsu11_sol_2c}
\eeq

The fermions of set (iia) transform, with respect to the subgroup 
(\ref{su2su9}), according to
\beq
3[5]_{11} = 3\{ (1,[\bar 4]_9) + (2,[4]_9) + (1,[3]_9)\}
\label{iiadec1}
\eeq
\beq
2[\bar 3]_{11} = 2 \{ (1,[\bar 3]_9) + (2,[\bar 2]_9) + (1,[\bar 1]_9)\}
\label{iiadec2}
\eeq
\beq
2[\bar 1]_{11} = 2 \{ (1,[\bar 1]_9) + (2,1) \} \ . 
\label{iidec3}
\eeq
With the SU(2)$_w$ couplings small, the nonsinglet SU(9)$_{SCC}$ fermion 
content is thus 
\beq
\{ f \}= 
4[\bar 1]_9 + 4[\bar 2]_9 + 3[3]_9 + 2[\bar 3]_9 + 6[4]_9 + 3[\bar 4]_9 \ . 
\label{iichiral}
\eeq
Hence, the SU(9)$_{SCC}$ sector is a chiral gauge theory.  If the
SU(9)$_{SCC}$ gauge interaction were asymptotically free and hence increased as
the energy scale decreased below $M_{GU}$, one could proceed to the next step
and analyze self-breaking condensate formation in the theory.  However, we find
that the SU(9)$_{SCC}$ interaction is non-asymptotically free, having a leading
coefficient of its beta function equal to 
\beq
b_0^{SU(9)_{SCC}} = -\frac{353}{3} \quad ({\rm class} \ iia) \ . 
\label{b0su9caseiia}
\eeq
With respect to the subgroup (\ref{su6su5}), the (left-handed chiral) fermions
of the set (iia) decompose according to
\beqs
3[5]_{11} & = & 3\{ (1,1) + ([1]_6, \bar 5) + ([2]_6,\overline{10}) +
([3]_6,10) \cr\cr
& + & ([\bar 2]_6,5) + ([\bar 1]_6,1) \} 
\label{3copiesof5_11decomp}
\eeqs
\beqs
2[8]_{11} \approx 2[\bar 3]_{11} & = & 2\{(1,10)+([\bar 1]_6,\overline{10})+
\cr\cr & + & ([\bar 2]_6,\bar 5) + ([\bar 3]_6,1) \}
\label{8_11_decomp}
\eeqs
and
\beq
2[10]_{11} \approx 2[\bar 1]_{11} = 2 \{ ([\bar1]_6,1))+(1,\bar 5) \} \ . 
\label{10_11_decomp}
\eeq
With the SU(5)$_{GU}$ couplings small, the nonsinglet left-handed fermions
transform according to the following SU(6)$_{SC}$ representations:
\beq
\{ f \} = 15[1]_6 + 25[\bar 1]_6 + 30 [2]_6 + 25 [\bar 2]_6 + 32 [3]_6 ] , 
\label{iiachiral}
\eeq
where we have used the fact that $[3]_6$ is equivalent to $[\bar 3]_6$.  Hence,
the SU(6)$_{SC}$ gauge interaction is chiral.  However, this class of 
models is disfavored because the SU(6)$_{SC}$ gauge interaction is not
asymptotically free; the leading coefficient of the beta function is 
\beq
b_0^{SU(6)_{SC}}=-\frac{386}{3} \quad ({\rm class} \ iia) \ . 
\label{b0su6caseiia}
\eeq
Hence, the SU(6)$_{SC}$ gauge coupling gets smaller rather than larger as the
energy scale decreases from high values, precluding the possibility of
condensate formation and self-breaking of SU(6)$_{SC}$ to extract the
SU(5)$_{TC}$ group and a $N_{g\ell}=1$ generation of SM fermions.

We next consider the subclass (iib).  The fact that an SU($N$) gauge theory
with odd $N \ge 5 $ and left-handed fermion content given by $n_i=0$ for
$i=1,3,...,N-2$ and $n_i=1$, $i=2,4,...,N-1$ is anomaly-free was shown in
\cite{g79}.  With respect to the subgroup (\ref{su2su9}), the fermions for this
class decompose according to
\beq
[2]_{11} = (1,[2]_9)+(2,[1]_9)+(1,1) 
\label{iibdec1}
\eeq
\beq
[4]_{11} = (1,[4]_9)+(2,[3]_9)+(1,[2]_9)
\label{iibdec2}
\eeq
\beq
[6]_{11} \approx [\bar 5]_{11} = (1,[4]_9)+(2,[\bar 4]_9)+(1,[\bar 3]_9) 
\label{iibdec3}
\eeq
\beq
[8]_{11} \approx [\bar 3]_{11} = (1,[\bar 3]_9)+(2,[\bar 2]_9)+(1,[\bar 1]_9)
\label{iibdec4}
\eeq
\beq
[10]_{11} \approx [\bar 1]_{11} = (1,[\bar 1]_9)+(2,1) \ . 
\label{iibdec5}
\eeq
With the SU(2)$_w$ couplings small, the nonsinglet SU(9)$_{SCC}$ fermion 
sector is then
\beqs
\{ f \} & = & 2\{ [1]_9 + [\bar 1]_9 + [2]_9 + [\bar 2]_9 + \cr\cr
& &     [3]_9 + [\bar 3]_9 + [4]_9 + [\bar 4]_9 \} \ . 
\label{iibsu9fermions}
\eeqs
Hence, although the SU(11) gauge interaction is chiral, the SU(9)$_{SCC}$ gauge
interaction is vectorial. Even if the SU(9)$_{SCC}$ interaction were
asymptotically free, this vectorial property would disfavor this class of
models because it would not self-break.  The SU(9)$_{SCC}$ interaction is
actually not asymptotically free; we calculate that
\beq
b_0^{SU(9)_{SCC}} = -\frac{157}{3} \quad ({\rm class} \ ii.b) \ . 
\label{b0su9caseiib}
\eeq

With respect to the subgroup (\ref{su6su5}), the fermion decompose according to
\beq
[2]_{11} = (1,10)+([1]_6,5)+([2]_6,1) 
\label{2_11_decomp}
\eeq
\beqs
[4]_{11} & = & (1,\bar 5)+([1]_6,\overline{10})+([2]_6,10) + \cr\cr
& + &  ([3]_6,5) + ([\bar 2]_5,1) 
\label{4_11_decomp}
\eeqs
and
\beqs
[6]_{11} & \approx & [\bar 5]_{11} = (1,1)+([\bar 1]_6,5)+
([\bar 2]_6,10)+ \cr\cr
& + & ([\bar 3]_6,\overline{10})+([2]_6,\bar 5)+([1]_6,1) \ . 
\label{6_11_decomp}
\eeqs
with the decompositions of $[8]_{11} \approx [\bar 3]_{11}$ and $[10]_{11}
\approx [\bar 1]_{11}$ given above. With the SU(5)$_{GU}$ couplings small, the
nonsinglet fermion content under SU(6)$_{SC}$ is 
\beq
16 \{ [1]_6 + [\bar 1]_6 + [2]_6 + [\bar 2]_6 + [3]_6 \} \ . 
\label{iibvectorial}
\eeq
We find that the SU(6)$_{SC}$ gauge interaction for this set is not
asymptotically free, with a leading coefficient of its beta function equal to
\beq
b_0^{SU(6)_{SC}} =-\frac{190}{3} \quad ({\rm class} \ iib) \ . 
\label{b0su6caseiib}
\eeq
This disfavors this class of models.

We have analyzed the class (iic) in a similar manner.  Decomposing the fermion
representations with respect to the subgroup (\ref{su2su9}) and cataloguing the
resultant SU(9)$_{SCC}$ content, we obtain the following nonsinglet
SU(9)$_{SCC}$ fermions:
\beq
\{ f \} = 4[1]_9 + 4[2]_9 + 4[3]_9 + 5[\bar 3]_9 + 7[4]_9 + 10[\bar 4]_9 \ . 
\label{iicfermionssu9}
\eeq
Hence, the SU(9)$_{SCC}$ gauge theory is chiral.  However, we find that the
SU(9)$_{SCC}$ gauge interaction is non-asymptotically free, with
\beq
b_0^{SU(9)_{SCC}} = -239 \quad ({\rm class} \ iic) \ . 
\label{iicb0su9}
\eeq
Decomposing the fermion representations with respect to the subgroup
(\ref{su6su5}), and cataloguing the resultant SU(6)$_{SC}$ content, we find
the SU(6)$_{SC}$ theory is chiral, but not asymptotically free, with 
\beq
b_0^{SU(6)_{SC}}=-250 \quad ({\rm class} \ iic) \ . 
\label{iicb0su6}
\eeq
For the same reasons as were given above, this model is thus disfavored as a
promising candidate for unification.

\section{$N_{SC}=7$, $G={\rm SU}(12)$}

We next study the case where $N_{SC}=7$, so that $N=N_{SC}+5=12$ and ${\bf n} =
(n_1,...,n_{11})_{12}$.  With $N_{gh}+N_{g\ell}=N_{gen.}=3$ and
$N_{SC}-N_{TC}=N_{g\ell}$, one has, {\it a priori}, four possibilities for the
manner in which the SM fermion generations arise, as specified by $(N_{gh},N_{g
\ell},N_{TC})$, namely (3,0,7), (2,1,6), (1,2,5), and (0,3,4).  The conditions
(\ref{no5}) and (\ref{no10bar}) that the theory should not contain any $5_L$ or
$\overline{10}_L$ yield
\beq
n_1=n_3=n_8=n_{10}=0 \ , 
\label{nzerosu12}
\eeq
and eq. (\ref{generationcondition}) is 
\beq
N_{gh}=n_2+n_9=n_4+n_{11} \ . 
\label{ngenhsu12}
\eeq
The condition of zero gauge anomaly, eq. (\ref{n.a}), is 
\beq
8n_2+48n_4+42(n_5-n_7)-27n_9-n_{11} = 0 \ . 
\label{anomalysu12}
\eeq
For a given value of $N_{gh}=3-N_{g \ell}$, these are three 
linear equations for the seven quantities $n_2$, $n_4$, $n_5$, $n_6$, $n_7$, 
$n_9$, and $n_{11}$. The solution implies the relations 
\beq
n_4 = \frac{1}{7}\Big [ 6(-n_5+n_7) + 5n_9 - N_{gh} \Big ]
\label{n4su12}
\eeq
and
\beq
n_{11} = \frac{1}{7}\Big [ 6(n_5-n_7) -5n_9 +8N_{gh} \Big ] \ . 
\label{n11su12}
\eeq

If $N_{gh}=0$, then $n_4=-n_{11}$, so the only allowed values are
$n_4=n_{11}=0$. It follows that $n_2=n_9=0$ also, and, substituting these
values into eqs. (\ref{n4su12}) and (\ref{n11su12}), one obtains $n_5=n_7$.
Thus, this class of solutions, which we denote as (i), has an ${\bf n}$ vector
equal to
\beq
{\bf n} = (0,0,0,0,s,t,s,0,0,0,0) \ , 
\label{su12ng0}
\eeq
where $s$ and $t$ are non-negative integers.  Since $[6]_{12} \approx [\bar
6]_{12}$ and $[5]_{12} \approx [\bar 7]_{12}$, this SU(12) theory is vectorial,
and hence so are resultant SU(10)$_{SCC}$ and SU(5)$_{SC}$ theories.  Hence,
even if the SCC and SC interactions were asymptotically free (which they are
not), these sectors would not self-break via condensate formation as would be
necessary in order to extract the TC theory and the SM fermion generations.  In
order to minimize the number of fermions in an effort to maintain asymptotic
freedom, we consider the two minimal classes (cases), (ia) $s=0$, $t=1$; and
(ib) $s=1$, $t=0$.  We find that
\beq
b_0^{SU(10)_{SCC}} = -\frac{142}{3} \quad ({\rm class} \ (ia))
\label{su12ia}
\eeq
and
\beq
b_0^{SU(10)_{SCC}} = -\frac{310}{3} \quad ({\rm class} \ (ib)) \ , 
\label{su12ib}
\eeq
which disfavors these cases from further consideration. 

Among other solutions, we focus on one that minimize the fermion content in an
effort to preserve asymptotic freedom.  We find cases with minimal 
${\bf n}$ vectors for $N_{gh}=3$.  Among these, the minimal one has 
\beq
(ii): \quad {\bf n} = (0,1,0,1,0,0,0,0,2,0,2) \ . 
\label{su12ii}
\eeq
We find that this yields a chiral SU(10)$_{SCC}$ gauge 
interaction, as desired, but the SU(10)$_{SCC}$ sector is not 
asymptotically free: 
\beq
b_0^{SU(10)_{SCC}} = -\frac{112}{3} \quad ({\rm class} \ (ii)) \ . 
\label{b0su12ii}
\eeq
We have found similar non-asymptotically free SCC sectors for other solutions
for this $N_g=3$ case, and also for cases with $N_g=1,2$. Our results suggest
that non-asymptotically free SCC and SC sectors appear to be a generic problem
with models having unification groups SU($N$) with $N \ge 11$.

\section{Discussion and Conclusions}

On the basis of our results, we can infer some generalizations concerning this
type of approach to unification of gauge symmetries in theories with dynamical
symmetry breaking.  We first recall some findings from Ref. \cite{tg} for
SU($N$) models with $N$ up to 10.  In that study, several cases were found that
satisfied the various necessary conditions listed above, including anomaly
cancellation, potential for $N_g=3$ standard-model fermion generations, absence
of SC-singlet fermions with exotic SM quantum numbers, etc., and for which the
$G_{SCC}$ gauge interaction was asymptotically free. However, in many of these
cases, this SCC gauge symmetry is vectorial, so that as the energy scale
decreases from $M_{GU}$, the SCC interaction eventually becomes strong,
confines, and produces a bilinear fermion condensate, but this condensate is
invariant under $G_{SCC}$, so this group does not self-break, as is necessary
to peel off the SC and color groups, and eventually the TC group.  One model
with $G={\rm SU}(10)$ and fermion content specified by ${\bf
n}=(0,0,0,1,0,0,1,0,0)_{10}$ yielded an asymptotically free chiral gauge sector
for $G_{SCC}$, but the condensate formation via the most attractive channel did
not produce an acceptable low-energy theory.  

   In the present work, we have searched for more promising models by examining
higher values of $N$, including $N=11$ and $N=12$.  Here we have encountered a
problem that was already present for a number of the models considered in
Ref. \cite{tg} with $N \le 10$, namely the property that the models contain
sufficiently many fermions that $G_{SCC}$ is not asymptotically free.  This
feature tends to preclude the desired scenario in which the SU($N_{SCC}$) group
would become strongly coupled as the energy scale decreases below $M_{GU}$ and
would self-break via formation of fermion condensates to separate out the
SU(3)$_c$ and SU($N_{SC}$) groups, and thus the SU($N_{TC}$) group.  This
appears to be a generic problem.  Thus, the necessary conditions stipulated
above, in their entirety, constitute a significant challenge for a viable
unification model.

Although our results are somewhat negative, the knowledge that we have gained
concerning models embodying the present type of approach is useful for
continuing efforts to construct theories that could unify the standard-model
gauge symmetries with gauge interactions that would become strong on the TeV
scale and cause dynamical electroweak symmetry breaking.  One may anticipate
that data from the CERN Large Hadron Collider, soon to go into operation, will
elucidate the question of the origin of electroweak symmetry breaking.  If
there is evidence that this symmetry breaking is dynamical, it will be
interesting to pursue further the goal of higher unifcation addressed here.

\begin{acknowledgments}

The present research was partially supported by the grant NSF-PHY-06-53342.  
R.S. thanks N. Christensen for collaboration on the earlier related work in
Ref. \cite{tg}.

\end{acknowledgments}

\begin{widetext} 

\begin{table}
\caption{\footnotesize{Some properties of the models discussed in the text with
$G_{SC}$ and $G_{SM}$ unified in a simple group $G$.  Here, $G_{SC}={\rm
SU}(N_{SC})$, $G_{TC}={\rm SU}(N_{TC})$, and $G_{SC} \supseteq G_{TC}$.  The
column marked ``SCC'' lists some properties of the SU($N_{SCC}$) theory
combining the ${\rm SU}(N_{SC})$ and SU(3)$_c$ groups.  See text for further
definitions and discussion.  The fermion content is indicated by the vector
${\bf n}$ (with subscript omitted for brevity).  The notation ``no sol.'' means
that (in the dynamical framework used) there is no solution to the requirements
of absence of any SU($N$) gauge anomaly, well-defined SM fermion generations,
and $N_{gen.}=3$. The notation VGT and CGT indicate that the gauge
interaction is vectorial and chiral, respectively; AF and NAF mean
asymptotically free and non asymptotically free, respectively. The $N_{(1,1)}$
is the number of electroweak-singlet neutrinos. The results up to $N=10$ from
\cite{tg} are included for comparative purposes.}}
\begin{center}
\begin{tabular}{|c|c|c|c|c|c|c|c|c|} \hline\hline
$N$ & $N_{SCC}$ & $N_{SC}$ & $N_{TC}$ & $N_{g\ell}$ & $N_{gh}$ &
${\bf n}$ & SCC & $N_{(1,1)}$ \\ \hline
7  & 5 & 2 & 2 & 0 & 3 & no sol.     & $-$        & $-$   \\ \hline
8  & 6 & 3 & 3 & 0 & 3 & (0200103)   & VGT, \, AF & 1     \\ \hline
8  & 6 & 3 & 2 & 1 & 2 & no sol.     & $-$        & $-$   \\ \hline
9  & 7 & 4 & 4 & 0 & 3 & no sol.     & $-$        & $-$   \\ \hline
9  & 7 & 4 & 3 & 1 & 2 & (01010101)  & VGT, \, AF & 1     \\ \hline
9  & 7 & 4 & 2 & 2 & 1 & no sol.     & $-$        & $-$   \\ \hline
10 & 8 & 5 & 5 & 0 & 3 & (000300300) & CGT, \, NAF& 0     \\ \hline
10 & 8 & 5 & 4 & 1 & 2 & (000200200) & CGT, \, NAF& 0     \\ \hline
10 & 8 & 5 & 3 & 2 & 1 & (000100100) & CGT, \, AF & 0     \\ \hline
10 & 8 & 5 & 3 & 2 & 1 & (000110100) & CGT, \, NAF& 2     \\ \hline
10 & 8 & 5 & 2 & 3 & 0 & (000010000) & VGT, \, AF & 2     \\ \hline
11 & 9 & 6 & 6 & 0 & 3 & no sol.     & $-$                 & $-$ \\ \hline
11 & 9 & 6 & 5 & 1 & 2 & (0000300202)& iia, \, CGT, \ NAF  & 3   \\ \hline
11 & 9 & 6 & 5 & 1 & 2 & (0101010101)& iib, \, VGT, \ NAF  & 1   \\ \hline
11 & 9 & 6 & 5 & 1 & 2 & (0202050000)& iic, \, CGT, \ NAF  & 5   \\ \hline
11 & 9 & 6 & 4 & 2 & 1 & no sol.     & $-$                 & $-$ \\ \hline
11 & 9 & 6 & 3 & 3 & 0 & (0000110000)& i, \,    VGT, \ NAF & 2   \\ \hline
12 &10 & 7 & 4 & 3 & 0 & (00000100000)& ia, \, VGT, \ NAF  & 0   \\ \hline
12 &10 & 7 & 4 & 3 & 0 & (00001010000)& ib, \, VGT, \ NAF  & 2   \\ \hline
12 &10 & 7 & 7 & 0 & 3 & (01010000202)& ii, \, CGT, \ NAF  & 0   \\ \hline
\hline
\end{tabular}
\end{center}
\label{properties}
\end{table}

\end{widetext}


\newpage

\begin{thebibliography}{99}

\bibitem{tc}
%
S. Weinberg, Phys. Rev. D {\bf 19}, 1277 (1979); 
L. Susskind, {\it ibid.} D {\bf 20}, 2619 (1979). 

\bibitem{etc}
S. Dimopoulos and L. Susskind, Nucl. Phys. B {\bf 155}, 237 (1979);
E. Eichten and K. Lane, Phys.  Lett. B {\bf 90}, 125 (1980). 

\bibitem{etcrev}
Some recent reviews are K. Lane, hep-ph/0202255; C. Hill and E. Simmons,
Phys. Rep. {\bf 381}, 235 (2003);
R. S. Chivukula, M. Narain, J. Womersley, in http://pdg.lbl.gov; and 
R. Shrock, in {\it The Origin of Mass and Strongly Coupled Gauge Theories - 
SCGT06}, Nagoya, eds. M. Harada, M. Tanabashi, and K. Yamawaki 
(World Scientific, Singapore, 2008).

\bibitem{swtc}
T. Appelquist and F. Sannino, Phys. Rev. D {\bf 59}, 067702 (1999); 
S. Ignjatovic, L. C. R. Wijewardhana, and T. Takeuchi, Phys. Rev.
D {\bf 61}, 056006 (2000); 
M. Harada, M. Kurachi, and K. Yamawaki, Prog. Theor. Phys. {\bf 115}, 765
(2006); 
M. Kurachi and R. Shrock, Phys. Rev. D {\bf 74}, 056003 (2006).

\bibitem{adss}
D. K. Hong and H.-U. Yee, Phys. Rev. D {\bf 74}, 015011 (2006);
J. Hirn and V. Sanz, Phys. Rev. Lett. {\bf 97}, 121803 (2006);
M. Piai, hep-ph/0608241; K. Agashe, C. Csaki, C. Grojean, and 
M. Reece, JHEP 0712:003,2007.

\bibitem{lrs}
T. Appelquist and R. Shrock, Phys. Rev. Lett. {\bf 90}, 201801 (2003).

\bibitem{fs}
E. Farhi and L. Susskind, Phys. Rev. D {\bf 20}, 3404 (1979).

\bibitem{tg}
N. D. Christensen and R. Shrock, Phys. Rev. D {\bf 72}, 035013 (2005).

\bibitem{gg}
H. Georgi and S. Glashow, Phys. Rev. Lett. {\bf 32}, 438 (1974).  We use only
the gauge and SM fermion content of this model, not the Higgs fields. 

\bibitem{hr}
N. D. Christensen and R. Shrock, Phys. Lett. B {\bf 632}, 92 (2006); 
S. B. Gudnason, T. A. Ryttov, and F. Sannino, Phys. Rev. D {\bf 76}, 
015005 (2007). 

\bibitem{ou}
A. Davidson, P. D. Mannheim, and K. C. Wali, Phys. Rev. D {\bf 26}, 1133
(1982); G. Grunberg, Phys. Rev. D {\bf 38}, 1012 (1988); J. D. Lykken and
S. Willenbrock, Phys. Rev. D {\bf 49}, 4902 (1994). 

\bibitem{nt}
T. Appelquist and R. Shrock, Phys. Lett. {\bf B548}, 204 (2002).

\bibitem{ckm}
T. Appelquist, M. Piai, and R. Shrock, Phys. Rev. D {\bf 69}, 015002 (2004).

\bibitem{fr}
P. Frampton, Phys. Rev. Lett. {\bf 43}, 1912 (1979); Erratum {\it ibid.},
{\bf 44}, 299 (1980).

\bibitem{rds}
S. Raby, S. Dimopoulos, and L. Susskind, Nucl. Phys. {\bf B} 169, 373 (1980).

\bibitem{g79}
 H. Georgi, Nucl. Phys. B {\bf 156}, 126 (1979). 

\end{thebibliography}
\end{document}